\documentstyle[aps,epsf]{revtex}  

\def\a{\alpha}

\def\g{\gamma}
\def\d{\delta}
\def\e{\epsilon}

\def\ro{\rho}
\def\si{\sigma}
\def\A{{\rm\scriptscriptstyle A}}
\def\H{{\rm\scriptscriptstyle H}}
\def\T{{\rm\scriptscriptstyle T}}
\def\HA{(\H,\A)}
\def\pt{p_\T}

\def\ga{g_\A}
\def\gh{g_\H}

\def\frac#1#2{{#1 \over #2}}
\def\slash#1{\rlap/#1}
\def\GeV{{\rm GeV}}

\def\la{\langle}
\def\ra{\rangle}
\def\l{\left}
\def\r{\right}
\def\M{{\cal M}}
\def\o{\over}
\def\beq{\begin{equation}}
\def\beqa{\begin{eqnarray}}
\def\eeq{\end{equation}}
\def\eeqa{\end{eqnarray}}

\def\tr{{\rm tr}}
\def\no{\nonumber}

\def\qq{\qquad}

\def\etal{{\it et al.}}
\begin{document}        

\baselineskip 14pt
\title{Amplitudes for Higgs Bosons plus Four Partons}
\author{Russel P. Kauffman}
\address{Department of Physics, Muhlenberg College, Allentown, PA 18104}
\author{Satish V. Desai} 
\address{Department of Physics, 
State University of New York at Stony Brook, Stony Brook, NY 11794} 
\author{Dipesh Risal} 
\address{Department of Biochemistry and Biophysics, University of Rochester,
Rochester, NY 14627}  
\maketitle              

\begin{abstract}        
In this talk we consider amplitudes for processes involving a Higgs boson, 
either scalar or pseudoscalar, plus four light partons.  These amplitudes are
relevant to the production of a Higgs boson plus two jets in hadronic 
collisions.  They are also relevant to calculating the transverse momentum 
spectrum for Higgs bosons at next-to-leading order in the strong coupling.
We work in the limit that the top quark is much
heavier than the Higgs bosons and use effective Lagrangians for the 
interactions of gluons with the Higgs bosons.  We present the amplitudes
involving a Higgs boson and: 1) four gluons; 2) two quarks and two 
gluons; and 3) four quarks.  We show that the pseudoscalar amplitudes are 
nearly identical to those for the scalar case, the only differences being 
the overall size and the relative signs between terms.  
\
\end{abstract}   	

\section{Introduction}               
Extensions of the Standard Model with enlarged Higgs sectors have richer 
particle content 
than the minimal Standard Model; in general, neutral pseudoscalars (with 
respect to their fermion couplings)
and charged scalars
as well as extra neutral scalars are present\cite{guide}. 
In this paper we will study
the amplitudes involving a neutral Higgs boson, scalar ($H$) or pseudoscalar 
($A$), and four light partons. This talk summarizes published work 
in Refs.~\cite{hgggg,hjj,ajj}.

We will present helicity amplitudes for the following processes:
\beqa
H(A) &\to& gggg \no \\
H(A) &\to& q\bar q gg\no \\
H(A) &\to& q\bar q q^{(\prime)} \bar q^{(\prime)}.
\eeqa
These amplitudes are a part of the calculation of the next-to-leading order 
transverse momentum distribution for Higgs bosons.  They can be combined with 
the virtual corrections to the production of a Higgs boson plus one jet to make
the total cross section.  The production of a Higgs boson plus two jets has 
interest in its own right because it provides information about the 
environment in which Higgs bosons are produced, allowing one to address the
question of how often there are extra jets in events with Higgs bosons.
Lastly, these amplitudes would also be a part of the calculation of the 
next-to-next-to-leading order cross section for Higgs boson production.

\section{Effective Lagrangians}
We will employ the approximation that the Higgs boson is much lighter than the 
top quark: $M \ll m_t$.  This greatly simplifies the calculation and is 
reasonably accurate even when $M \sim m_t$.  Furthermore, the shape of the
Higgs-boson $\pt$ distribution at low $\pt$ 
is independent of $m_t$ and so the
approximate results give some information about heavy Higgs bosons as well.
We use the following effective Lagrangians\cite{rusk,aonejet} to couple the 
Higgs bosons to gluons:
\beqa
{\cal L}_{\H}&=&-{1\over 4} \gh G_{\mu \nu}^a G_{\mu \nu}^a H , \no\\
{\cal L}_{\A} &=& \ga G_{\mu\nu}^a \tilde{G}_{\mu\nu}^a A,
\eeqa
where $G^a_{\mu \nu}$ is the field strength of the SU(3) color
gluon field, 
$\tilde{G}_{\mu\nu}^a$ is its dual, 
$\tilde{G}_{\mu\nu}^a = {1\o2}\e^{\mu\nu\rho\sigma} G_{\rho\sigma}^a$,
and $H,A$ are the Higgs fields.  We choose the 
couplings between
the top quark and the Higgs scalar and pseudoscalar to be $m_t/v$ and 
$m_t \g_5 /v$, respectively,
where $v$ is the vacuum expectation
value parameter, $v^2=(G_F\sqrt{2})^{-1}=(246~\GeV)^2$.
The effective couplings to gluons are then
given by  $\gh = \a_s /(3 \pi v)$ and $\ga = \a_s /(8 \pi v)$. 

The effective Lagrangians generate vertices involving the Higgs bosons
and two, three, or four gluons.   We assign outgoing gluon momenta and spin 
indices $p_1^\mu, p_2^\nu, p_3^\rho, p_4^\sigma$
and color indices $a,b,c,d$.  Labeling the vertex involving a Higgs scalar
and $n$ gluons as $V_n^\H$ we have 
\beqa
V_2^\H &=& i\gh \d^{ab} (g^{\mu\nu}p_1 \cdot p_2 - p_1^\nu p_2^\mu), \no \\
V_3^\H &=& -g\gh f^{abc}\l [(p_1-p_2)^\ro g^{\mu\nu}  
                             + (p_2-p_3)^\mu g^{\nu\ro}
                             + (p_3-p_1)^\nu g^{\ro\mu} \r],    \no \\
V_4^\H &=& -i g^2 \gh \l[
f_{abe}f_{cde}( g^{\mu\ro}g^{\nu\si} - g^{\mu\si}g^{\nu\ro} )
+f_{ace}f_{bde}( g^{\mu\nu}g^{\ro\si} - g^{\mu\si}g^{\nu\ro} ) 
+f_{ade}f_{bce}( g^{\mu\nu}g^{\ro\si} - g^{\mu\ro}g^{\nu\si} ). \r]
\eeqa
In similar notation, the vertices for the Higgs pseudoscalar are
\beqa
V_2^\A  &=& -i \ga \delta^{ab} \e^{\mu\nu\rho\sigma}
                               p_{1\rho} p_{2\sigma},\no \\
V_3^\A  &=& -g\ga f^{abc} \e^{\mu\nu\rho\sigma}
                               (p_1 + p_2 + p_3)_\sigma.
\label{avert}
\eeqa
The four-gluon pseudoscalar vertex vanishes as it 
is proportional to the completely 
antisymmetric combination of structure constants:
\beq
f^{abe}f^{cde} - f^{ace}f^{bde} + f^{ade}f^{bce} = 
-2 \tr \{ [T^a,T^b][T^c,T^d] - [T^a,T^c][T^b,T^d] + [T^a,T^d][T^b,T^c] \} = 0,
\eeq
where the $T^i$ are the SU(3) generators.

\section{Spinor-Product Formalism}
We will compute helicity amplitudes using a spinor-product 
formalism\cite{helic,ber}.  We refer the reader to Ref. \cite{hjj}
for details of our implementation.  
We will use the convention that all the 
particles are outgoing. The momenta
of the massless particles are labeled $p_1,~p_2,~p_3,~p_4$ and the momentum of
the Higgs boson is labeled $p$.
We will use the shorthand notations $\la ij \ra$ to denote a left-handed 
spinor with momentum $p_i$ contracted on the left with a right-handed spinor
with momentum $p_j$ and $[ij]$ to represent such a contraction with the
right-handed spinor on the left.  We also define the invariant masses
$S_{ij} = (p_i+p_j)^2$   and $S_{ijk} = (p_i+p_j+ p_k)^2$.

Amplitudes for processes involving the Higgs pseudoscalar contain expressions
of the form
$\e_{\mu\nu\rho\sigma}w^\mu x^\nu y^\ro z^\si$ where $w,x,y,z$ are
momenta, polarization vectors, or fermion currents.  These contractions can
be written in terms of spinor products through the following procedure
\cite{ajj}.
We first write
\beq
\e_{\mu\nu\rho\sigma}w^\mu x^\nu y^\ro z^\si =
{1\o4i} \tr\{\slash w \slash x \slash y \slash z \g_5\}  
= {1\o4i} \tr\{\slash w \slash x \slash y \slash z (P_+ - P_-)\},
\label{eq:econtract}  
\eeq
where the projection operators are $P_\pm = (1\pm\g_5)/2$.
Each slashed vector can be written in terms of outer products of spinors:
\beq
\slash w = w_+|w_1{+} \ra \la w_2{+}| + w_-  |w_3{-} \ra \la w_4{-}|.
\label{eq:decompose}
\eeq
Inserting Eq.~(\ref{eq:decompose}) into Eq.~(\ref{eq:econtract})
and expressing the trace in terms of matrix multiplication, we have
\beq
\e_{\mu\nu\rho\sigma}w^\mu x^\nu y^\ro z^\si = {1\o4i} ( w_+ \la w_2{+}| \slash x \slash y \slash z | w_1{+} \ra
          - w_- \la w_4{-}| \slash x \slash y \slash z | w_3{-} \ra ),
\eeq
which reduces to spinor products upon application of Eq.~(\ref{eq:decompose})
to $\slash x$, $\slash y$, and $\slash z$.
  
\section{Limits}

One of the virtues of computing helicity amplitudes in the spinor product
formalism is that their limiting behavior is easily made manifest.
We know on general grounds that in the limit that an external parton becomes 
soft or two particles become collinear the Higgs boson plus four parton
amplitudes must factorize into a sum of eikonal factors times amplitudes for
Higgs bosons plus three partons\cite{parki}.  
Additionally, we know that the amplitudes
for Higgs scalars going to three partons are identical, up to couplings and
phases, to those for Higgs pseudoscalars\cite{aonejet,ajj}.

The amplitudes also have specific limits when the four-momentum of the Higgs
boson goes to zero.  The scalar amplitudes become proportional to the pure
QCD amplitudes for four partons.  The pseudoscalar amplitudes, on the other
hand, must vanish.  This behavior is due to the coupling of the pseudoscalar
field to $G\tilde G$, which is a total derivative.  Inspection of the vertices
involving the pseudoscalar, Eq.~(\ref{avert}), shows that they are 
proportional to the pseudoscalar momentum: 
$V_2^{\A} \sim \e^{\mu\nu\rho\si}k_{1\rho}p_\si$ and 
$V_3^{\A} \sim \e^{\mu\nu\rho\si}p_\si$.

\section{Results}
The four-gluon amplitudes are written in the dual-color decomposition
\cite{parki,parkii,parkiii}.
The scattering amplitude for a Higgs boson and four gluons with external
momenta $p_1$,...,$p_4$, colors $a_1$,...,$a_4$, and
helicities $\lambda_1$,...,$\lambda_4$ is 
\beq
\M_{\HA}= 2 g_{\HA} g^2
\sum_{\rm perms} {\rm tr} (T^{a_1}...T^{a_4})m(p_1,\epsilon_1;
...;p_4,\epsilon_4),
\label{eq:dual}
\eeq
where the sum is over the non-cyclic permutions of the
momenta.  For the helicity choices ++++ and ${-}$+++ the subamplitudes for
Higgs pseudoscalars are identical to their scalar counterparts and
can be obtained from the following\cite{hgggg,hjj,ajj}:
\beqa
&&m(1^+,2^+,3^+,4^+)={M^4\over \la 1 2\ra \la 2 3\ra
   \la 3 4\ra \la 4 1\ra} 
\label{eq:pppp}\\ 
&&m(1^-,2^+,3^+,4^+) = -
{\la 1{-}|\slash p |3{-} \ra^2 [24]^2 \over S_{124} S_{12} S_{14}}     
-{\la 1{-}|\slash p | 4{-} \ra^2 [2 3]^2 \over S_{123} S_{12} S_{23}} 
-{\la 1{-}|\slash p | 2{-} \ra^2 [3 4]^2 \over S_{134} S_{14} S_{34}} 
\no \\
&&\qq\qq\qq\qq
 +{[2 4] \over 
[ 1 2 ] \la 2 3\ra \la 3 4 \ra   [ 4 1]} 
\l\{ S_{23} {\la 1{-} |\slash p | 2{-} \ra \over \la 4 1\ra } 
+       S_{34} {\la 1{-} |\slash p | 4{-} \ra \over \la 1 2\ra }
-[2 4 ] S_{234}\r\}. 
\label{eq:mppp}
\eeqa
The subamplitudes for the helicity choice ${-}{-}$++ differ between 
scalar and pseudoscalar by a relative sign between terms.  All the 
subamplitudes for this case can be obtained from 
\beq
m_{\HA}(1^-,2^-,3^+,4^+)=-{\la 1 2\ra^4\over \la 12 \ra 
\la 23 \ra \la 34\ra \la 41\ra}
  \mp {[34]^4 \over [1 2] [23] [34] [41]} ,
\label{eq:mmpp}
\eeq
where the upper sign goes with the $H$ and the lower with the $A$.

As was the case for the four-gluon amplitudes, 
the calculation of the $q \bar q gg$ amplitudes 
can be simplified by a judicious choice of color decomposition
\cite{kunszt,parkii}.
We assign momentum labels 1,2,3,4 and color labels $i,j,a,b$ to the quark, 
anti-quark and two gluons, respectively.  The amplitude is then written
\beq
\M_{\HA} = -i g^2 g_{\HA}
\sum_{\rm perms} (T^a T^b)_{ij} m(p_3,\e_3;p_4,\e_4),
\label{eq:qdual}
\eeq
where the sum runs over the two gluon permutations.
We label the helicity amplitudes by the helicity of the quark, anti-quark
and the two gluons (in that order).  The quark and anti-quark
always have opposite
helicities.  In the case where the gluons have identical helicities  we find 
that the scalar and pseudoscalar amplitudes are identical\cite{hjj,ajj}: 
\beq
m^{+-++}(3,4) = { \la 2{-}|\slash p| 3{-} \ra^2 \o S_{124} }
                 { [14] \o \la 24 \ra } \l({1\o S_{12}} + {1\o S_{14}} \r)
            -{ \la 2{-}|\slash p| 4{-} \ra^2 \o S_{123} S_{12}}
             { [13] \o \la 23 \ra }
            +{ \la 2{-}|\slash p| 1{-} \ra^2 \o [12] 
            \la 23 \ra \la 24 \ra \la 34 \ra }.
\label{eq:qpmpp}
\eeq
To get the subamplitude with the other ordering, $m^{+-++}(4,3)$, exchange
$p_3 \leftrightarrow p_4$ in this expression.  
When the two gluons have opposite helicities the scalar and pseudoscalar
amplitudes differ by a relative sign between terms:
\beqa
m_{\HA}^{+-+-}(3,4) &=& { [13]^3 \o [12][14][34] }
 \mp {\la24\ra^3 \o \la12\ra \la23\ra \la34\ra } 
\label{eq:qpmpmi} \\
m_{\HA}^{+-+-}(4,3) &=& -{ [13]^2 [23] \o [12][24][34] }
                  \pm {\la14\ra \la24\ra^2 \o \la12\ra \la13\ra \la34\ra } ,
\label{eq:qpmpmii} 
\eeqa
where the upper signs go with the $H$ and the lower with the $A$.
The rest of the subamplitudes can be obtained by parity, charge conjugation
and Bose symmetry transformations.

In the case of the four-quark amplitude with two different quark pairs
the helicity of each quark must be opposite to its anti-quark partner.  
We choose momentum labels 1,2,3,4 and color labels $i,j,k,l$ for the first
quark, first anti-quark, second quark and second anti-quark respectively and 
label the helicities in that order.
All the helicity 
amplitudes (including those for identical quark pairs) can be obtained from
\cite{hjj,ajj} 
\beq
\M_{\HA}^{+-+-} = i g_{\HA} g^2 T^a_{ij}T^a_{kl} 
\l( {\la24\ra^2 \o \la12\ra \la34\ra} \pm { [13]^2 \o [12][34] } \r),
\label{eq:aqqqq}
\eeq
where the upper sign is for the $H$ and the lower for the $A$.

\section{Conclusion}
The following pattern emerges from the amplitudes presented in the previous
section: those amplitudes which violate helicity are identical for the scalar
and pseudoscalar, modulo the different couplings;  those amplitudes which 
conserve helicity differ for the scalar and pseudoscalar by a relative sign
between two terms.  This allows the soft and collinear limits to be the same
for the scalar and pseudoscalar, modulo phases.

We see that the helicity-violating amplitudes vanish in the limit that the 
four-momentum of the Higgs boson goes to zero.  This is consistent with the
general result for the Higgs scalar:  the scalar amplitudes reduce to pure QCD
amplitudes in this limit and pure QCD conserves helicity.  It is also 
consistent with the general result for the pseudoscalar which is that all
the amplitudes must vanish when the pseudoscalar momentum goes to zero.
Furthermore, we see that the helicity-conserving amplitudes for the 
pseudoscalar also vanish in this limit.  The two terms in each expression 
cancel (whereas for the scalar case they add).


\begin{references}  
%
\bibitem{guide}
J.~F.~Gunion, H.~E.~Haber, G.~L.~Kane, and S.~Dawson, {\it The Higgs
Hunter's Guide} (Addison-Wesley, Menlo Park, California, 1990), p.~201.
%
\bibitem{hgggg} S.~Dawson and R.~P.~Kauffman, Phys.~Rev.~Lett.
{\bf 68} 2273 (1992).
%
\bibitem{hjj} R.~P.~Kauffman, S.~V.~Desai and D.~Risal, Phys.~Rev.~D 
{\bf 55}, 4005 (1997).
%
\bibitem{ajj} R.~P.~Kauffman, S.~V.~Desai, Phys.~Rev.~D 
{\bf 59}, 057504 (1999).
%
\bibitem{rusk} A.~Vainshtein, M.~Voloshin, V.~Zakharov, and
M.~Shifman, Sov.~J.~Nucl.~Phys. {\bf 30} 429 (1979).
%
\bibitem{aonejet} R.~P.~Kauffman and W.~Schaffer, Phys.~Rev.~D {\bf 49},
551 (1994);
M.~Spira \etal, Nucl.~Phys.~{\bf B453}, 17 (1995).
%
\bibitem{helic} Z.~Xu,
D.~Zhang, and L.~Chang, Nucl.~Phys.~{\bf B291}
392 (1987).
\bibitem{ber} F.~Berends {\it et al.}, Nucl.~Phys.~{\bf B206} 61 (1982); 
{\it ibid.}~{\bf B239} 382 (1984);
{\it ibid.}~{\bf B239} 395 (1984); {\it ibid.}~{\bf B254} 265 (1986).
%
\bibitem{parki}
M.~Mangano, S.~Parke, and Z. Xu, {\it Nucl. Phys.} {\bf B298} (1988) 653.
%
\bibitem{parkii}
M.~Mangano and S.~Parke, {\it Nucl. Phys.} {\bf B299} (1988) 673.
%
\bibitem{parkiii}
M.~Mangano and S.~Parke, {\it Phys. Reps.} {\bf 200} (1991) 301.
%
\bibitem{kunszt}
Z. Kunszt, {\it Nucl. Phys.} {\bf B271} (1986) 333.
\end{references}
\end{document}